\documentclass [12pt]{article}

\usepackage{amsmath,amssymb,amsthm,epsfig}
\def\lanbox{\hbox{$\, \vrule height 0.25cm width 0.25cm depth 0.01cm \,$}}

\usepackage{color}

\def\ulambda{{\underline{\lambda}}}
\def\Xom{{\mathcal L}^2_{\ulambda}(\Omega)}

\def\Zom{{\mathcal H}^{1}_\ulambda(\Omega)}
\def\dYom{\dot{\mathcal H}^{1/2}_\ulambda(\Omega)}

\def\Zoms{{\mathcal H}^{s}_\ulambda(\Omega)}
\def\dZoms{\dot{\mathcal H}^{s}_\ulambda(\Omega)}

\def\N{{\mathbb N}}

\def\eqnn{\begin{eqnarray*}}
\def\eeqnn{\end{eqnarray*}}
\def\eqn{\begin{eqnarray}}
\def\eeqn{\end{eqnarray}}

\def\prf{\begin{proof}}
\def\endprf{\end{proof}}

\theoremstyle{plain}

\numberwithin{equation}{section}

\begin{document}

\title{Existence and nonlinear stability of stationary states for the 
semi-relativistic Schr\"odinger-Poisson system}

\date{}

\maketitle

\centerline{Walid Abou Salem}

\centerline{Department of Mathematics and Statistics, 
University of Saskatchewan,}

\centerline{Saskatoon, SK, S7N 5E6, Canada}
 
\centerline{e-mail: walid.abousalem@usask.ca}

\bigskip

\centerline{Thomas Chen}

\centerline{Department of Mathematics, University of
Texas at Austin,}

\centerline{Austin, TX, 78712, USA}

\centerline{e-mail: tc@math.utexas.edu}
 
\bigskip

\centerline{Vitali Vougalter}

\centerline{University of Cape Town, Department of Mathematics
and Applied Mathematics,}
 
\centerline{Private Bag, Rondebosch 7701, South Africa}

\centerline{e-mail: Vitali.Vougalter@uct.ac.za}

\medskip

\bigskip
\bigskip
\bigskip

\noindent {\bf Abstract.}  We study the stationary states of the semi-relativistic Schr\"odinger-Poisson system in the repulsive (plasma physics) Coulomb case. In particular, we establish the existence and the nonlinear stability of a wide class of stationary states by means of the energy-Casimir method. We generalize
the global well-posedness result of ~\cite{ACV11} for the semi-relativistic
Schr\"odinger-Poisson system to spaces with higher regularity.

\bigskip
\bigskip

\noindent {\bf Keywords:} relativistic kinetic energy; Schr\"odinger-Poisson
system; Hartree-von Neumann equation; stationary solutions; nonlinear stability;
global existence and uniqueness

\bigskip

\noindent {\bf AMS subject classification:} 82D10, 82C10   

\newpage

\section{Introduction}

\bigskip
 
In this work, we prove the existence and the nonlinear stability for a certain
class of stationary solutions of the semi-relativistic Schr\"odinger-Poisson 
system in a finite volume domain with   Dirichlet boundary 
conditions. Such a system describes the mean-field dynamics of semi-
relativistic quantum particles (for instance, in the
case of heated plasma), with the particles moving with extremely high 
velocities. Let us consider semi-relativistic quantum particles confined in a
domain $\Omega\subset {\mathbb R}^{3}$ which is an open set with a ${C}^{2}$ 
boundary and $|\Omega|<\infty$. These particles are interacting by means of the
electrostatic field they collectively generate. In the mean-field limit, the 
density matrix describing the {\it mixed} state of the system solves the 
Hartree-von Neumann equation
\begin{equation}
\label{HVN}
\begin{cases}
i \partial_{t}{\rho(t)} = [H_V,\rho(t)], \ \ x\in \Omega, \ \ t
\geq 0\\
-\Delta V = n(t,x), \ \ n(t,x)=\rho(t,x,x),
\ \rho(0)=\rho_0
\end{cases},
\end{equation}
with  Dirichlet boundary conditions, $\rho(t,x,y)=0$ if $x$ or 
$y\in\partial\Omega$, for $t\geq0$. The single particle Hamiltonian is given by 
\begin{equation}
\label{HV}
H_{V}:=T_{m}+V(t,x).
\end{equation}
The relativistic kinetic energy operator $T_{m}:=\sqrt{-\Delta+m^{2}}-m$ is 
 defined by means of the spectral calculus. 
In system (\ref{HVN}) and further below, $\Delta$ stands for the Dirichlet 
Laplacian on $L^{2}(\Omega)$, and $m>0$ is the single particle mass. We refer 
to ~\cite{A10} and ~\cite{AS09} for a derivation of the analogous system of 
equations in the {\it non-relativistic} case. Due to the fact that
$\rho(t)$ is a nonnegative, self-adjoint and trace-class operator acting on 
$L^2(\Omega)$, we can expand its kernel, for every $t\in\mathbb{R}_+$,  with 
respect to an orthonormal basis of $L^2(\Omega)$. 
We denote this kernel at the initial time $t=0$ by $\rho_0,$ 
\begin{equation}
\label{kernelinit}
\rho_0(x,y) = \sum_{k\in {\mathbb N}} \lambda_k \psi_k(x)\overline{\psi_k(y)}.
\end{equation}
Here $\{\psi_k\}_{k\in {\mathbb N}}$ stands for an orthonormal basis of 
$L^2(\Omega)$, such that $\psi_k|_{\partial\Omega}=0$ for all $k\in\N$, and the
coefficients are given by
\begin{equation}
\label{lambda}
\ulambda:=\{\lambda_{k}\}_{k\in {\mathbb N}}\in l^{1}, \quad \lambda_{k}\geq 0, \quad
\sum_{k\in {\mathbb N}}\lambda_{k}=1.
\end{equation} 
In \cite{ACV11}, we showed that there exists a one-parameter family of 
complete orthonormal bases of $L^2(\Omega)$, $\{\psi_k(t)\}_{k\in {\mathbb N}}$, 
with $\psi_k(t)|_{\partial\Omega}=0$ for all $k\in\N$, and for 
$t\in\mathbb{R}_+$,  
such that the kernel of the density matrix $\rho(t)$, which satisfies system
 (\ref{HVN}), can be expressed as 
\begin{equation}
\label{kernel}
\rho(t,x,y) = \sum_{k\in {\mathbb N}} \lambda_k \psi_k(t,x)\overline{\psi_k(t,y)}.
\end{equation}
As a consequence of the particular commutator structure of \eqref{HVN}
(where $\rho(t)$ and $-iH_V$ satisfy the conditions of a Lax pair),
the corresponding flow of $\rho(t)$ leaves its spectrum invariant.
Accordingly, the coefficients $\ulambda$ are {\em independent} of $t$.
This isospectrality is crucial for the stability analysis for stationary states based
on the Casimir energy method employed in this paper; see also \cite{A69,G99,IZL94,MRW02,R99}.

By substituting the expression (\ref{kernel}) in system (\ref{HVN}), one can verify that the  
one-parameter family of orthonormal vectors  $\{\psi_k(t)\}_{k\in {\mathbb N}}$ 
satisfies the semi-relativistic Schr\"odinger-Poisson system equivalent to
(\ref{HVN}) and given by
\begin{equation}
\label{sch}
i\frac{\partial \psi_{k}}{\partial t}=T_{m}\psi_{k}+V\psi_{k}, \quad k\in 
{\mathbb N},
\end{equation}  
\begin{equation}
\label{p}
-\Delta V[\Psi]=n[\Psi], \quad \Psi:=\{\psi_{k}\}_{k=1}^{\infty},
\end{equation} 
\begin{equation} 
\label{n}
n[\Psi(t,x)]=\sum_{k=1}^{\infty}\lambda_{k}|\psi_{k}|^{2},
\end{equation} 
with initial data $\{\psi_k(0)\}_{k=1}^\infty$. Here, the potential function 
$V[\Psi]$  is the solution of the Poisson equation (\ref{p}). Both $V[\Psi]$ 
and $\psi_k(t)$, for all $k\in\N$, satisfy the Dirichlet boundary conditions
\begin{equation}
\label{dV}
\psi_k(t,x) \; , \; \;
V(t,x)=0, \ t\geq 0, \ \forall x\in \partial \Omega.
\end{equation}
The global well posedness for system (\ref{sch})-(\ref{dV}) was established in
the recent work ~\cite{ACV11}. 
Analogous results were derived before in 
the nonrelativistic case in a finite volume domain with Dirichlet boundary
conditions in ~\cite{BM91}, and in the whole space of ${\mathbb R}^{3}$ in
~\cite{BM91} and  ~\cite{IZL94}. 

In this paper, we are interested in the properties of {\it stationary states} which 
occur when $\rho(t)=f(H_{V})$ for some function $f$.
Substituting the latter in (\ref{HVN}),  
the commutator on the right side of the first equation of system (\ref{HVN}) vanishes, and the 
density matrix is time independent. The precise properties of the 
distribution function $f$ will be discussed below. 
The solution of the Schr\"odinger-Poisson system corresponding to the stationary states is 
$$ 
\psi_{k}(t,x)=e^{-i\mu_{k}t}\psi_{k}(x), \quad k\in {\mathbb N},
$$
such that the potential function $V[\Psi]$ is time independent, 
$\mu_{k}\in {\mathbb R}$ are the eigenvalues of the Hamiltonian (\ref{HV})
and $\psi_{k}(x)$ are the corresponding eigenfunctions. 

The organization of this article is as follows.
 In Section 2, we describe the class of stationary states we will study, and state our hypotheses and main results about nonlinear stability and existence of stationary states. In Section 3, we derive some preliminary results. In Section 4, we prove the nonlinear stability of the stationary states of the
semi-relativistic Schr\"odinger-Poisson system via the energy-Casimir 
functional as a Lyapunov function (see the statement of Theorem 1). In Section
5 we define the dual functional and in Section \ref{sec:ExStatStates} study its 
properties using the methods of convex analysis, and show that it admits a 
unique maximizer (see Theorem 2), which implies the existence of a stationary 
state for our Schr\"odinger-Poisson system. In an Appendix, we discuss generalizing the well-posedness result of \cite{ACV11} to spaces with higher regularity.

\bigskip



\setcounter{equation}{0}

\section{The Model and Statement of the Main Results}

\bigskip

The state space for the Schr\"odinger-Poisson system  is defined as
$$
{\cal L}:=\{ (\Psi, \ulambda) \ | \ \Psi=\{\psi_{k}\}_{k=1}^{\infty}\subset
H_{0}^{\frac{1}{2}}(\Omega)\cap H^{1}(\Omega) \; \; is \ a \ complete \ 
orthonormal \ system \ in \ L^{2}(\Omega), 
$$
$$
\ulambda=\{\lambda_{k}\}_{k=1}^{\infty}\in {l}^1, \quad \lambda_{k}\geq 0, \ k\in
{\mathbb N}, \quad \sum_{k=1}^{\infty}\lambda_{k}\int_{\Omega}|\nabla \psi_{k}|^{2}
dx<\infty \},
$$
see ~\cite{ACV11}.

In order to precisely define the class of stationary states we will study, we need to introduce the {\it Casimir class} of functions. We say that a function $f: {\mathbb R}\to {\mathbb R}$ is of Casimir class ${\cal C}$ 
if and only if it possesses the following properties:

\bigskip

(i) $f$ is continuous, such that $f(s)>0$ for $s\leq s_{0}$ and $f(s)=0$ when
$s\geq s_{0},$ with some $s_{0}\in ]0, \infty]$,

\bigskip

(ii) $f$ is strictly decreasing on $]-\infty, s_{0}]$, such that
$\hbox{lim}_{s\to -\infty}f(s)=\infty$,

\bigskip

(iii) there exist constants $\varepsilon>0$ and $C>0,$ such that for $s\geq 0$
the estimate
\begin{equation}
\label{fu}
f(s)\leq C(1+s)^{-5-\varepsilon}
\end{equation}
holds.

Note that $s_0$ acts as an {\it ``ultra-violet"} cut-off, and we can take is as large as we wish.

Consider the quadruple $(\Psi_{0}, \ulambda_{0}, \mu_{0}, V_{0})$ with
$(\Psi_{0}, \ulambda_{0})\in {\cal L}$, 
$\mu_{0}=\{\mu_{0,k}\}_{k=1}^{\infty}$ real valued, and the potential function 
$V_{0}\in H_{0}^{1}(\Omega)\cap H^{2}(\Omega),$ such that the stationary 
Schr\"odinger-Poisson system holds
\begin{equation}
\label{sch0}
(T_{m}+V_{0})\psi_{0,k}=\mu_{0,k}\psi_{0,k}, \quad k\in {\mathbb N},
\end{equation}
\begin{equation}
\label{p0}
-\Delta V_{0}=n_{0}=\sum_{k=1}^{\infty}\lambda_{0,k}|\psi_{0,k}|^{2} \,,
\end{equation}
with 
\begin{equation}
\label{lf0}
\lambda_{0,k}=f(\mu_{0,k}), \quad k\in {\mathbb N},
\end{equation}
where $f\in {\cal C}$. Then, the corresponding density matrix 
$\rho_{0}=f(T_{m}+V_{0})$ solves the stationary state Hartree-von
Neumann equation
$$
[H_{V_{0}},\rho_{0}]=0. 
$$

\bigskip

{\bf Remark.} In the nonrelativistic case, the Casimir class was defined 
similarly in ~\cite{MRW02} with the exception that the rate of decay of the
distribution function $f$ was assumed to be smaller. A good example of 
$f\in {\cal C}$ is the function decaying exponentially as $s\to \infty$
with the cut-off level $s_{0}=\infty$. This is the Boltzmann distribution
$f(s):=e^{-\beta s}, \ \beta>0$.

\bigskip

In order to prove the nonlinear stability of the stationary states, we will use the energy-Casimir method. This method was used in \cite{A69} for fluid problems, and in \cite{G99,R99} for studying stationary states of kinetic equations, in particular, Vlasov-Poisson systems.
Here, we extend the energy-Casimir functional used in \cite{MRW02} to the semi-relativistic case.
For $f\in {\cal C}$, let us define
\begin{equation}
\label{F}
F(s):=\int_{s}^{\infty}f(\sigma)d\sigma, \quad s\in {\mathbb R}.
\end{equation}
Note that the function defined via (\ref{F}) is decreasing, continuously
differentiable, nonnegative and is strictly convex on its support. Moreover,
for $s\geq 0$
\begin{equation}
\label{Fu}
F(s)\leq C(1+s)^{-4-\varepsilon}.
\end{equation}
Its Legendre
(Fenchel) transform is given by
\begin{equation}
\label{LF}
F^{*}(s):=\hbox{sup}_{\lambda\in {\mathbb R}}(\lambda s-F(\lambda)), \quad s\leq 0.
\end{equation}
We define the energy-Casimir functional for a given $f$ as
\begin{equation}
\label{HC}
{\cal H}_{C}(\Psi, \ulambda):=\sum_{k=1}^{\infty}F^{*}(-\lambda_{k})+
{\cal H}(\Psi, \ulambda), \quad (\Psi, \ulambda)\in {\cal L}.
\end{equation}
In particular,
${\cal H}_{C}$
is conserved along solutions of the Schr\"odinger-Poisson system,
as a consequence of isospectrality of the flow of $\rho(t)$,  which is equivalent to the $t$-independence of $\lambda_k$.
Our main results in this paper address the existence and stability of stationary states 
that are given by (\ref{sch0})-(\ref{lf0}), for $f\in {\cal C}.$ Stability is controlled by the following main theorem.


\bigskip

{\bf Theorem 1.} {\it Let $(\Psi_{0}, \ulambda_{0}, \mu_{0}, V_{0})$ be a 
stationary state of the semi-relativistic Schr\"odinger-Poisson system, where
$$
\lambda_{0,k}=f(\mu_{0,k}), \quad {k\in \mathbb N}
$$
with some $f\in {\cal C}$ and $(\Psi_{0}, \ulambda_{0})\in {\cal L}$. Let 
$(\Psi(t), \ulambda)$ be a solution of the Schr\"odinger-Poisson system, such
that initial datum $(\Psi(0), \ulambda)\in {\cal L}$. Then, for all $t\geq 0$,
the estimate
$$
\frac{1}{2}\|n[\Psi(t),\ulambda]-n_{0}\|_{\dot H^{-1}(\Omega)}^{2}\leq
{\cal H}_{C}(\Psi(0), \ulambda)-{\cal H}_{C}(\Psi_{0}, \ulambda_{0})
$$
holds, such that the stationary state is nonlinearly stable. Here, $\dot{H}^{-1}(\Omega)$ is the dual of $\dot{H}^1(\Omega)$ with norm $\|u\|_{\dot{H}^{-1}(\Omega)} = (u, (-\Delta)^{-1}u)^{1/2}_{L^2(\Omega)}.$}

\bigskip

To prove the existence of stationary states,
we   introduce the dual of the energy-Casimir functional. 
To this end, we let, for $\Lambda>0$ fixed,   
$$
{\cal G}(\Psi, \ulambda,V, \sigma):=\sum_{k=1}^{\infty}[F^{*}(-\lambda_{k})+
\lambda_{k}\int_{\Omega}[|T_{m}^{\frac{1}{2}}\psi_{k}|^{2}+V|\psi_{k}|^{2}]dx]-
\frac{1}{2}\int_{\Omega}|\nabla V|^{2}dx+\sigma \Big[\sum_{k=1}^{\infty}\lambda_{k}-
\Lambda \Big].
$$
Here, $\sigma\in {\mathbb R}$ is a Lagrange multiplier. 

The dual functional to ${\cal H}_{C}$ is given by 
\begin{equation}
\label{fi}
\Phi(V,\sigma):=\hbox{inf}_{\Psi, \ulambda}{\cal G}(\Psi,\ulambda, V, \sigma).
\end{equation}
The infimum in the formula above is taken over all $\ulambda\in l_{+}^{1}$ and
all complete orthonormal sequences $\Psi$ from $L^{2}(\Omega)$.
Let us consider only non-negative potential
functions and define
$$
H_{0,+}^{1}(\Omega):=\{V\in H_{0}^{1}(\Omega) \ | \ V\geq 0 \}.
$$
The following is our   main result about the existence of stationary states.

\bigskip

{\bf Theorem 2.} {\it Let $f\in {\cal C}$ and $\Lambda>0$ be fixed. The
functional $\Phi$}
$$
(V,\sigma)\in H_{0,+}^{1}(\Omega)\times {\mathbb R} \to -\frac{1}{2}\int_{\Omega}
|\nabla V|^{2}dx-\hbox{Tr}[F(T_{m}+V+\sigma)]-\sigma \Lambda
$$
{\it is continuous, strictly concave, bounded from above and $-\Phi(V,\sigma)$ 
is coercive. There exists a unique maximizer $(V_{0},\sigma_{0})$ of 
$\Phi(V,\sigma)$. Let 
$\{\psi_{0,k}\}_{k=1}^{\infty}$ be the orthonormal sequence of eigenfunctions of
the Hamiltonian $T_{m}+V_{0}$ corresponding to the eigenvalues 
$\{\mu_{0,k}\}_{k=1}^{\infty}$ and $\lambda_{0,k}:=f(\mu_{0,k}+\sigma_{0})$. Then
$(\Psi_{0}, \ulambda_{0}, \mu_{0}, V_{0})$ is a stationary state of the semi-relativistic 
Schr\"odinger-Poisson system, where 
$\sum_{k=1}^{\infty}\lambda_{0,k}=\Lambda$ and 
$(\Psi_{0}, \ulambda_{0})\in {\cal L}$. }

\bigskip

We will prove Theorem 1 in Section 4, and Theorem 2 in Section 5.

\bigskip


\setcounter{equation}{0}

\section{Preliminaries}

\bigskip

We have the following elementary lemma.

\bigskip

{\bf Lemma 3.} {\it For $(\Psi, \ulambda)\in {\cal L}$ we have 
$$
n_{\psi,\lambda}:=\sum_{k\in {\mathbb N}}\lambda_{k}|\psi_{k}|^{2}\in L^{2}(\Omega).
$$
Let $V_{\psi,\lambda}$ stand for the Coulomb potential induced by $n_{\psi,\lambda}$,
such that
$$
-\Delta V_{\psi,\lambda}(x)=n_{\psi,\lambda}(x), \quad x\in \Omega; \quad
V_{\psi,\lambda}(x)=0, \quad x\in \partial \Omega.
$$ 
Then} $V_{\psi,\lambda}\in H_{0}^{1}(\Omega)\cap H^{2}(\Omega)$.

\bigskip

{\it Proof.} We easily express the norm as
$$
\|n_{\psi,\lambda}\|_ {L^{2}(\Omega)}^{2}=\sum_{k,s=1}^{\infty}\lambda_{k}\lambda_{s}
(|\psi_{k}|^{2}, |\psi_{s}|^{2})_ {L^{2}(\Omega)}.
$$
Here and further below, the inner product of two functions 
$f(x),g(x)\in L^{2}(\Omega)$ is denoted as 
$(f,g)_ {L^{2}(\Omega)}:=\int_{\Omega}f(x)\bar{g}(x)dx$.
Application of the Schwarz inequality to the right side of the identity above
yields the upper bound
$$
\Bigg(\sum_{k=1}^{\infty}\lambda_{k}\sqrt{\int_{\Omega}|\psi_{k}|^{4}dx}\Bigg)^{2},
$$
which can be estimated from above by applying the Schwarz inequality as well.
Thus, we obtain 
$$
|\Omega|^{\frac{1}{3}}\Bigg(\sum_{k=1}^{\infty}\lambda_{k}\Bigg(\int_{\Omega}
|\psi_{k}|^{6}dx \Bigg)^{\frac{1}{3}}\Bigg)^{2}.
$$
We next make use of the Sobolev inequality
\begin{equation}
\label{sineq}
\int_{\Omega}|\nabla f|^{2}dx\geq c_{s}\Bigg(\int_{\Omega}|f|^{6}dx\Bigg)^{\frac{1}
{3}}, 
\end{equation}
in which the constant $c_{s}$ is given on p.186 of ~\cite{LL97}. Noting that a
function compactly supported in the set $\Omega$ can be extended by zero
to the whole space of ${\mathbb R}^{3}$, we arrive at the upper bound
$$
\frac{|\Omega|^{\frac{1}{3}}}{{c_{s}}^{2}}\Bigg(\sum_{k=1}^{\infty}\lambda_{k}
\int_{\Omega}|\nabla \psi_{k}|^{2}dx\Bigg)^{2}<\infty
$$
by means of the definition of the state space ${\cal L}$ given above, such 
that $n_{\psi, \lambda}\in L^{2}(\Omega)$. Note that the particle density
$n_{\psi, \lambda}$ vanishes on the boundary of the set $\Omega$ by means of
formula (\ref{n}) and boundary conditions (\ref{dV}). Therefore, 
$\Delta V_{\psi, \lambda}\in L^{2}(\Omega)$. Let $\{\mu_{k}^{0}\}_{k\in {\mathbb N}}$
denote the eigenvalues of the Dirichlet Laplacian on $L^{2}(\Omega)$ and
$\mu_{1}^{0}$ is the lowest one of them. Note that
$$
\mu_{k}^{0}>0, \quad k\in {\mathbb N}.
$$
Since
$$
V_{\psi, \lambda}=(-\Delta)^{-1}n_{\psi, \lambda}, 
$$
we have that $\displaystyle{\|V_{\psi, \lambda}\|_{ L^{2}(\Omega)}\leq \frac{1}
{\mu_{1}^{0}}\|n_{\psi, \lambda}\|_{ L^{2}(\Omega)}<\infty}.$ Furthermore, 
$V_{\psi, \lambda}$ vanishes on the boundary of the set $\Omega$ via (\ref{dV}).
\hfill\lanbox

\bigskip

According to Theorem 1 of ~\cite{ACV11}, for every initial state 
$(\Psi(0), \ulambda)\in {\cal L}$, there exists a unique strong solution of
system (\ref{sch})-(\ref{dV}), where $(\Psi(t), \ulambda)\in {\cal L}$ for
all $t\geq 0$. Let us define the energy of a state 
$(\Psi, \ulambda)\in {\cal L}$ as
$$
{\cal H}(\Psi,\ulambda):=\sum_{k=1}^{\infty}\lambda_{k}\int_{\Omega}
|T_{m}^{\frac{1}{2}}\psi_{k}|^{2}dx+\frac{1}{2}\int_{\Omega}n_{\psi, \lambda}
V_{\psi, \lambda}dx=
$$
$$
=\sum_{k=1}^{\infty}\lambda_{k}\int_{\Omega}|T_{m}^{\frac{1}{2}}\psi_{k}|^{2}dx+
\frac{1}{2}\int_{\Omega}|\nabla V_{\psi, \lambda}|^{2}dx,
$$
which is a conserved quantity along solutions of the Schr\"odinger-Poisson
system (see Lemma 7 of ~\cite{ACV11}). 


Analogously to ~\cite{ACV11} we assume that $\lambda_{k}>0$ via density
arguments. To prove the nonlinear stability for a specified stationary state,
We have the following auxiliary statements.

\bigskip

{\bf Lemma 4.} {\it Let $f\in {\cal C}$.

\bigskip

a) For every $\beta>1$ there exists $C=C(\beta)\in {\mathbb R}$, such that
for $s\leq 0$ we have
$$
F(s)\geq -\beta s+C
$$
b) Let $V\in H_{0}^{1}(\Omega)$ and $V(x)\geq 0$ for $x\in \Omega$. Then
both operators $f(T_{m}+V)$ and $F(T_{m}+V)$ are trace class.}

\bigskip

{\it Proof.} The part a) of the lemma comes from the fact that function
$F(s)$ is smooth with the slope varying from $-\infty$ to $0$, and convex;
therefore, its graph is located above a tangent line to it.

For the Dirichlet eigenvalues of $-\Delta$ on 
$L^{2}(\Omega), \ \Omega\subset {\mathbb R}^{3}$, we will make use
of the semiclassical lower bound
$$
\mu_{k}^{0}\geq C k^{\frac{2}{3}}, \ k\in {\mathbb N}
$$
with a constant dependent on $|\Omega|<\infty$ (see e.g. ~\cite{LY83}).
Since the potential function $V(x)$ is nonnegative in $\Omega$ by assumption, 
we easily estimate from below the eigenvalues $\mu_{k}$ of the Hamiltonian
$T_{m}+V$ for  $k\in {\mathbb N}$ as
\begin{equation}
\label{mkl}
\mu_{k}\geq \sqrt{\mu_{k}^{0}+m^{2}}-m\geq (Ck^{\frac{1}{3}}-m)_+
\end{equation}
with the right side of the inequality above positive for $k$ large enough. For
the sharp semiclassical bounds on the moments of Dirichlet eigenvalues to
fractional powers see ~\cite{V11}. We express
$$
\hbox{Tr}F(T_{m}+V)=\sum_{k=1}^{\infty}F(\mu_{k})<\infty,
$$
since $F(s)$ is decreasing, satisfies estimate (\ref{Fu}), and the series with
a general term $(1+(Ck^{\frac{1}{3}}-m)_+)^{-4-\varepsilon}$ converges. Similarly,
$$
\hbox{Tr}f(T_{m}+V)=\sum_{k=1}^{\infty}f(\mu_{k})<\infty,
$$
due to the fact that $f(s)$ decreases, obeys bound (\ref{fu}) and the 
series with a general term $(1+Ck^{\frac{1}{3}}-m)^{-5-\varepsilon}$ is convergent.
This completes the proof of the part b) of the lemma. \hfill\lanbox

\bigskip

{\bf Lemma 5.} {\it Let $\psi\in H_{0}^{\frac{1}{2}}(\Omega)\cap H^{1}(\Omega)$ 
with $\|\psi\|_{L^{2}(\Omega)}=1$, the potential function $V\in H_{0}^{1}(\Omega)$
and $V(x)\geq 0$ for $x\in \Omega$. Then,
\begin{equation}
\label{Fi}
F(\langle \psi, (T_{m}+V)\psi \rangle)\leq \langle \psi, F(T_{m}+V)\psi \rangle
\end{equation}
holds with equality if $\psi$ is an eigenstate of the Hamiltonian $T_{m}+V$.
}

\bigskip

{\it Proof.} By means of the Spectral Theorem we have 
$$
T_{m}+V=\sum_{k=1}^{\infty}\mu_{k}P_{k},
$$
where the operators $\{P_{k}\}_{k=1}^{\infty}$ are the orthogonal projections
onto the bound states corresponding to the eigenvalues
$\{\mu_{k}\}_{k=1}^{\infty}$. Hence
$$
F(\langle \psi, (T_{m}+V) \psi \rangle)=F\Bigg(\sum_{k=1}^{\infty}\mu_{k}\|P_{k}
\psi\|_{L^{2}(\Omega)}^{2}\Bigg).
$$
The right side of (\ref{Fi}) can be easily written as 
$$
\sum_{k=1}^{\infty}F(\mu_{k})\|P_{k}\psi\|_{L^{2}(\Omega)}^{2}.
$$
Estimate (\ref{Fi}) follows from Jensen's inequality. When
$\psi$ is an eigenstate of the operator $T_{m}+V$ corresponding to an 
eigenvalue $\mu_{k}$, for some $k\in {\mathbb N}$, both sides of (\ref{Fi}) are equal to
$F(\mu_{k})$. Note that the converse of this statement does not hold in general.
Indeed, let us consider as $\psi$ a linear combination of more than one
eigenstate of the Hamiltonian with corresponding eigenvalues $\mu_{k}$ located
outside the support of $F(s)$. Then both sides of (\ref{Fi}) will be equal to
zero. \hfill\lanbox

\bigskip
The lemma below shows that a stationary solutions belong to the state space for the Schr\"odinger-Poisson system. 

\bigskip

{\bf Lemma 6.} {\it Let the quadruple $(\Psi_{0}, \ulambda_{0}, \mu_{0}, V_{0})$
satisfy equations (\ref{sch0}), (\ref{p0}) and (\ref{lf0}), where $\Psi_{0}$ is
a complete orthonormal system in $L^{2}(\Omega)$ and the distribution 
$f\in {\cal C}$. Then,
$$
\sum_{k=1}^{\infty}\lambda_{0,k}\int_{\Omega}|\nabla \psi_{0,k}|^{2}dx<\infty
$$ 
holds, such that $(\Psi_{0}, \ulambda_{0})\in {\cal L}$.}

\bigskip

{\it Proof.} We express the following quantity using (\ref{sch0}) and 
(\ref{lf0}) as
$$
\sum_{k=1}^{\infty}\lambda_{0,k}\|T_{m}^{\frac{1}{2}}\psi_{0,k}\|_{L^{2}(\Omega)}^{2}+
\int_{\Omega}|\nabla V_{0}|^{2}dx=
$$
\begin{equation}
\label{fmu}
=\sum_{k=1}^{\infty}\lambda_{0,k}((T_{m}+V_{0})
\psi_{0,k}, \psi_{0,k})_{L^{2}(\Omega)}=\sum_{k=1}^{\infty}f(\mu_{0,k})\mu_{0,k}.
\end{equation}
The potential function $V_{0}(x)\geq 0$ in $\Omega$ since it is superharmonic 
by means of (\ref{p0}), and vanishes on the boundary of $\Omega$. Thus, 
$\mu_{0,k}>0, \ k\in {\mathbb N}$ and via (\ref{fu}) the right side of 
(\ref{fmu}) can be bounded above by 
$$
\sum_{k=1}^{\infty}C(1+\mu_{0,k})^{-5-\varepsilon}\mu_{0,k}<\infty,
$$
which follows from the eigenvalue estimate (\ref{mkl}). We have also obtained
\begin{equation}
\label{TmV0}
\nabla V_{0}\in L^{2}(\Omega), \quad  T_{m}^{\frac{1}{2}}\psi_{0,k}\in 
L^{2}(\Omega), \ k\in {\mathbb N}.
\end{equation}
In fact, from equation (\ref{sch0}) with a nonnegative potential, we easily 
conclude that 
\begin{equation}
\label{Tmm0}
\|T_{m}^{\frac{1}{2}}\psi_{0,k}\|_{L^{2}(\Omega)}^{2}\leq \mu_{0,k}, \ k\in {\mathbb N}.
\end{equation}
Note that the standard requirement $V_{0}\in L^{1}(\Omega)$ (see e.g. p.234, 245
of ~\cite{LL97}) is satisfied here as well.
Throughout the article the operator $|p|:=\sqrt{-\Delta}$, which is the 
massless relativistic kinetic energy operator defined via the spectral 
calculus. Clearly, in the sense of quadratic forms, we have
\begin{equation}
\label{Tmp}
T_{m}\geq |p|-m,
\end{equation} 
such that (\ref{TmV0}) yields
$$
\||p|^{\frac{1}{2}}\psi_{0,k} \|_{L^{2}(\Omega)}^{2}\leq ((T_{m}+m)\psi_{0,k}, 
\psi_{0,k})_{L^{2}(\Omega)}=m+\|T_{m}^{\frac{1}{2}}\psi_{0,k}\|_{L^{2}(\Omega)}^{2}<\infty.
$$
Thus, $\psi_{0,k}\in H_{0}^{\frac{1}{2}}(\Omega), \ k\in {\mathbb N}$. Moreover,
let us make use of the relativistic Sobolev inequality (see e.g. p.183 of
~\cite{LL97}) for a function compactly supported in $\Omega$
\begin{equation}
\label{sr}
(f,|p|f)_{L^{2}(\Omega)}\geq c_{s}^{r}\|f\|_{L^{3}(\Omega)}^{2},
\end{equation}
which gives us $\psi_{0,k}\in L^{3}(\Omega), \ k\in {\mathbb N}$. 
Hence by means of H\"older's inequality, we arrive at
$$
\int_{\Omega}|V_{0}|^{2}|\psi_{0,k}|^{2}dx\leq \Bigg(\int_{\Omega}|V_{0}|^{6}dx\Bigg)^
{\frac{1}{3}}\Bigg(\int_{\Omega}|\psi_{0,k}|^{3}dx\Bigg)^{\frac{2}{3}}<\infty.
$$
Indeed, $V_{0}(x)\in L^{6}(\Omega)$ due to (\ref{TmV0}) along with (\ref{sineq}).
Therefore, $V_{0}\psi_{0,k}\in L^{2}(\Omega), \ k\in {\mathbb N}$. From equation
(\ref{sch0}) we easily deduce that 
$T_{m}\psi_{0,k}\in L^{2}(\Omega), \ k\in {\mathbb N}$ as well. Then the identity
\begin{equation}
\label{Tm12}
\|T_{m}\psi_{0,k}\|_{L^{2}(\Omega)}^{2}=\int_{\Omega}|\nabla \psi_{0,k}|^{2}dx-2m
\|T_{m}^{\frac{1}{2}}\psi_{0,k}\|_{L^{2}(\Omega)}^{2}
\end{equation}
via (\ref{TmV0}) yields $\nabla \psi_{0,k}\in L^{2}(\Omega)$ and 
$\psi_{0,k}\in H^{1}(\Omega), \ k\in {\mathbb N}$. By means of (\ref{lf0}), we 
have $\lambda_{0,k}\geq 0, \ k\in {\mathbb N}$. Convergence of the series on
the right side of (\ref{fmu}) implies
\begin{equation}
\label{lfm0}
\sum_{k=1}^{\infty}\lambda_{0,k}=\sum_{k=1}^{\infty}f(\mu_{0,k})<\infty,
\end{equation}
such that $\ulambda_{0}=\{\lambda_{0, k}\}_{k=1}^{\infty}\in l^{1}$. Let us make
use of identity (\ref{Tm12}) along with (\ref{Tmm0}), such that
$$
\sum_{k=1}^{\infty}\lambda_{0, k}\int_{\Omega}|\nabla \psi_{0,k}|^{2}dx=
\sum_{k=1}^{\infty}\lambda_{0, k} \{2m \|T_{m}^{\frac{1}{2}}\psi_{0,k}\|_{L^{2}(\Omega)}^{2}
+\|T_{m}\psi_{0,k}\|_{L^{2}(\Omega)}^{2}\}\leq
$$
\begin{equation}
\label{Tmub}
\leq 2m \sum_{k=1}^{\infty}f(\mu_{0,k})\mu_{0,k}+\sum_{k=1}^{\infty}\lambda_{0, k}
\|T_{m}\psi_{0,k}\|_{L^{2}(\Omega)}^{2}.
\end{equation}
The first term in the right side of (\ref{Tmub}) is finite as it was shown 
above. The second expression on the right side of the inequality above
can be written via (\ref{sch0}) as
\begin{equation}
\label{ttmV}
\sum_{k=1}^{\infty}\lambda_{0,k}\|\mu_{0,k}\psi_{0,k}-V_{0}\psi_{0,k}\|_{L^{2}(\Omega)}^{2}
=\sum_{k=1}^{\infty}\lambda_{0,k} \{\mu_{0,k}^{2}+\|V_{0}\psi_{0,k}\|
_{L^{2}(\Omega)}^{2}-2\mu_{0,k}\int_{\Omega}V_{0}|\psi_{0,k}|^{2}dx\}.
\end{equation}
Our goal is to prove that (\ref{ttmV}) is convergent. Indeed, (\ref{fu})
implies 
$$
\sum_{k=1}^{\infty}\lambda_{0,k} \mu_{0,k}^{2}\leq C \sum_{k=1}^{\infty}
(1+\mu_{0,k})^{-5-\varepsilon}\mu_{0,k}^{2}<\infty,
$$
due to the eigenvalue bound (\ref{mkl}). We estimate the second term on 
the right side of (\ref{ttmV}) using H\"older's inequality, such that
$$
\sum_{k=1}^{\infty}\lambda_{0,k}\|V_{0}\psi_{0,k}\|_{L^{2}(\Omega)}^{2}\leq 
\Bigg(\int_{\Omega}|V_{0}|^{6}dx\Bigg)^{\frac{1}{3}}\sum_{k=1}^{\infty}\lambda_{0,k}
\Bigg(\int_{\Omega}|\psi_{0,k}|^{3}dx \Bigg)^{\frac{2}{3}},
$$
where $V_{0}(x)\in L^{6}(\Omega)$ as discussed above. Let us make use of 
inequalities (\ref{sr}) and (\ref{Tmp}), such that
$$
\sum_{k=1}^{\infty}\lambda_{0,k}\|\psi_{0,k}\|_{L^{3}(\Omega)}^{2}\leq 
\frac{1}{c_{s}^{r}}\sum_{k=1}^{\infty}\lambda_{0,k}(|p|\psi_{0,k},\psi_{0,k})_
{L^{2}(\Omega)}\leq \frac{1}{c_{s}^{r}}\sum_{k=1}^{\infty}\lambda_{0,k}\{m+
\|T_{m}^{\frac{1}{2}}\psi_{0,k}\|_{L^{2}(\Omega)}^{2}\}<\infty
$$
due to estimates (\ref{lfm0}) and (\ref{fmu}). The last term in the right side
of (\ref{ttmV}) can be bounded above in the absolute value by applying the
Schwarz inequality to it twice, such that
$$
\sum_{k=1}^{\infty}\lambda_{0,k}\mu_{0,k}\|V_{0}\psi_{0,k}\|_{L^{2}(\Omega)}\leq
\sqrt{\sum_{k=1}^{\infty}\lambda_{0,k}\mu_{0,k}^{2}}\sqrt{\sum_{s=1}^{\infty}
\lambda_{0,s}\|V_{0}\psi_{0,s}\|_{L^{2}(\Omega)}^{2}}<\infty
$$
as it was shown above. \hfill\lanbox

\bigskip

{\bf Remark.} In the stationary situation, our semi-relativistic 
Schr\"odinger-Poisson problem can be easily written as
$$
-\Delta V_{0}=f(T_{m}+V_{0})(x,x), \quad x\in \Omega,
$$
$$
V_{0}(x)=0, \quad x\in \partial \Omega.
$$
Let us turn our attention to defining the corresponding Casimir functional
for a fixed $f\in {\cal C}$. 
The following elementary lemma gives the alternative representation
for the Legendre transform of our integrated distribution. Note that
$f\in {\cal C}$ considered on the $(-\infty, s_{0}]$ semi-axis has an inverse
$f^{-1}$.

\bigskip

{\bf Lemma 7.} {\it For the function $F(s)$ defined in (\ref{F}) and 
$s\leq 0$ we have}
\begin{equation}
\label{Fni}
F^{*}(s)=\int_{-s}^{0}f^{-1}(\sigma)d\sigma.
\end{equation}
{\it Proof.} Let us define
$$ 
g(\lambda):=\lambda s-F(\lambda), \quad \lambda \in {\mathbb R}, \quad s\leq 0,
$$
such that via (\ref{F}) we have $g'(\lambda)=s+f(\lambda)$. Hence, the maximal
value in the right side of (\ref{LF}) is attained at 
$\lambda^{*}:=f^{-1}(-s)$ and is equal to
$$
\varphi(s):=g(\lambda^{*})=f^{-1}(-s)s-\int_{f^{-1}(-s)}^{\infty}f(\sigma)d\sigma
$$
with $\varphi(0)=0$. Although $f$ is continuous and not necessarily differentiable, we can approximate it by a differentiable function. Let $f_\epsilon(s) =\frac{1}{2\epsilon} \int_{s-\epsilon}^{s+\epsilon}f(t)dt,$ and 
$$
\varphi_\epsilon(s)=f_\epsilon^{-1}(-s)s-\int_{f_\epsilon^{-1}(-s)}^{\infty}f_\epsilon(\sigma)d\sigma.
$$
A direct computation using the formula for the 
derivative of the inverse yields $\varphi_\epsilon'(s)=f_\epsilon^{-1}(-s)$. Integrating and taking the $\epsilon\rightarrow 0$ limit yields $\varphi(s)=\int_{-s}^{0}f^{-1}(\sigma)d\sigma$. \hfill\lanbox

\bigskip

In the next section, we prove the nonlinear stability of   
stationary states, using the energy-Casimir functional defined above.

\bigskip


\setcounter{equation}{0}

\section{Stability of stationary states}

\bigskip

In this section, we prove Theorem 1, which yields lower bound in terms of the 
electrostatic field. The auxiliary statement below is crucial for establishing this nonlinear
stability result.

\bigskip

{\bf Lemma 8.} {\it Let $V\in H_{0}^{1}(\Omega)$ and $V\geq 0$. 
(i) Then, for 
$(\Psi, \ulambda)\in {\cal L}$, the lower bound}
\begin{equation}
\label{l7in}
\sum_{k=1}^{\infty}\Bigg\{ F^{*}(-\lambda_{k})+\lambda_{k}\int_{\Omega}
[|T_{m}^{\frac{1}{2}}\psi_{k}|^{2}+V|\psi_{k}|^{2}]dx \Bigg \}\geq 
-\hbox{Tr}[F(T_{m}+V)].
\end{equation}
(ii) {\it Equality is attained for 
$(\Psi, \ulambda)=(\Psi_{V}, \ulambda_{V})$, where 
${\psi_{V,}}_{k}\in H_{0}^{\frac{1}{2}}(\Omega)\cap H^{1}(\Omega), \ k\in 
{\mathbb N}$ stands for the orthonormal sequence of eigenfunctions of the
Hamiltonian $T_{m}+V$ with corresponding eigenvalues ${\mu_{V,}}_{k}$ and
${\lambda_{V,}}_{k}=f({\mu_{V,}}_{k}), \ k\in {\mathbb N}$.}

\bigskip

{\it Proof.} According to definition (\ref{LF}), we have
$$
F^{*}(s)\geq \mu s-F(\mu), \quad \mu \in {\mathbb R}, \quad s\leq 0,
$$
which easily implies
\begin{equation}
\label{Flmk}
F^{*}(-\lambda_{k})+\lambda_{k}\mu_{k}\geq -F(\mu_{k}), \quad k\in {\mathbb N}.
\end{equation}
Now let
$$ 
\mu_{k}:=\int_{\Omega}\Bigg\{|T_{m}^{\frac{1}{2}}\psi_{k}|^{2}+V|\psi_{k}|^{2}\Bigg\}
dx=\langle \psi_{k},(T_{m}+V)\psi_{k}\rangle, \quad k\in {\mathbb N},
$$
which proves part $(i)$.
To prove part $(ii)$, we note that after summation, we arrive at
$$
\sum_{k=1}^{\infty}\Bigg\{F^{*}(-\lambda_{k})+\lambda_{k}\int_{\Omega}\Bigg\{
|T_{m}^{\frac{1}{2}}\psi_{k}|^{2}+V|\psi_{k}|^{2}\Bigg\}dx\Bigg\}\geq -
\sum_{k=1}^{\infty}F(\langle \psi_{k}, (T_{m}+V)\psi_{k}\rangle ). 
$$
Lemma 5 along with the definition of trace yields the lower bound for the
right side of the inequality above as
$$
-\sum_{k=1}^{\infty}\langle \psi_{k}, F(T_{m}+V)\psi_{k}\rangle=
-\hbox{Tr}(F(T_{m}+V)).
$$
Suppose that $(\Psi,\ulambda)=(\Psi_{V},\ulambda_{V})$, where 
${\psi_{V,}}_{k}$ are eigenfunctions of the Hamiltonian $T_{m}+V$ and $\mu_{k}$,
which are defined above are the corresponding eigenvalues 
${\mu_{V,}}_{k}, \ k\in {\mathbb N}$. Therefore, on the right side of the lower 
bound (\ref{l7in}) we have
$$
-\hbox{Tr}(F(T_{m}+V))=-\sum_{k=1}^{\infty}F({\mu_{V,}}_{k}).
$$
Next, we use the identity ${\lambda_{V,}}_{k}=f({\mu_{V,}}_{k})=-F'({\mu_{V,}}_{k})$.
Then, via Lemma 7, 
${F^{*}}'(-{\lambda_{V,}}_{k})=f^{-1}({\lambda_{V,}}_{k})={\mu_{V,}}_{k}, \ k\in
{\mathbb N}$. Using the argument of Lemma 7, we arrive at
$$
F^{*}(-{\lambda_{V,}}_{k})=\hbox{sup}_{\lambda\in {\mathbb R}}(-\lambda 
{\lambda_{V,}}_{k}-F(\lambda))=-f^{-1}({\lambda_{V,}}_{k}){\lambda_{V,}}_{k}
-F(f^{-1}({\lambda_{V,}}_{k}))=-{\lambda_{V,}}_{k}{\mu_{V,}}_{k}-F({\mu_{V,}}_{k}).
$$
Therefore, the left side of (\ref{l7in}) will be equal to 
$-\sum_{k=1}^{\infty}F({\mu_{V,}}_{k})$ as well. \hfill\lanbox

\bigskip

Armed with the technical lemma above, we may now prove our first main
statement.

\bigskip

{\it Proof of Theorem 2.} Let $(\Psi, \ulambda)\in {\cal L}$ and the potential
function $V=V_{\psi,\lambda}$ is induced by this state. Then we will use the 
following identity for the energy of the electrostatic field
$$
\frac{1}{2}\|n-n_{0}\|_{\dot H^{-1}(\Omega)}^{2}=
\frac{1}{2}\|\nabla V-\nabla V_{0}\|_{L^{2}(\Omega)}^{2}=\frac{1}{2}\int_{\Omega}
|\nabla V|^{2}dx+\frac{1}{2}\int_{\Omega}|\nabla V_{0}|^{2}dx+
\int_{\Omega}V_{0}\Delta V dx.
$$
By the definition of the energy-Casimir functional, this can be written as
$$
{\cal H}_{C}(\Psi, \ulambda)-\Big\{\sum_{k=1}^{\infty}\Big(F^{*}(-\lambda_{k})+
\lambda_{k}\int_{\Omega}|T_{m}^{\frac{1}{2}}\psi_{k}|^{2}dx \Big)-\frac{1}{2}
\int_{\Omega}|\nabla V_{0}|^{2}dx-\int_{\Omega}V_{0}\Delta V dx \Big\},
$$
which is equal to
$$
{\cal H}_{C}(\Psi, \ulambda)-\Big\{\sum_{k=1}^{\infty}\Big[F^{*}(-\lambda_{k})+
\lambda_{k}\int_{\Omega}(|T_{m}^{\frac{1}{2}}\psi_{k}|^{2}+V_{0}|\psi_{k}|^{2})dx \Big]
-\frac{1}{2}\int_{\Omega}|\nabla V_{0}|^{2}dx \Big\}.
$$
Applying first Lemma 8$(i)$, and subsequently Lemma 8$(ii)$, we obtain that the
expression above is bounded from above by 
$$
{\cal H}_{C}(\Psi, \ulambda)-\Big\{-\hbox{Tr}[F(T_{m}+V_{0})]-\frac{1}{2}
\int_{\Omega}|\nabla V_{0}|^{2}dx \Big\}=
$$
$$
={\cal H}_{C}(\Psi, \ulambda)-\Big\{\sum_{k=1}^{\infty}\Big[F^{*}(-\lambda_{0,k})+
\lambda_{0,k}\int_{\Omega}(|T_{m}^{\frac{1}{2}}\psi_{0,k}|^{2}+V_{0}|\psi_{0,k}|^{2})dx 
\Big]-\frac{1}{2}\int_{\Omega}|\nabla V_{0}|^{2}dx \Big\}=
$$
$$
={\cal H}_{C}(\Psi, \ulambda)-\Big\{\sum_{k=1}^{\infty}\Big[F^{*}(-\lambda_{0,k})+
\lambda_{0,k}\int_{\Omega}|T_{m}^{\frac{1}{2}}\psi_{0,k}|^{2}dx \Big]+
\frac{1}{2}\int_{\Omega}|\nabla V_{0}|^{2}dx \Big\}= 
$$
$$
={\cal H}_{C}(\Psi, \ulambda)-{\cal H}_{C}(\Psi_{0}, \ulambda_{0}).
$$
Since the Casimir functional is constant along the solutions of the 
Schr\"odinger-Poisson system, which is globally well-posed
(see ~\cite{ACV11}), for an initial condition 
$(\Psi(0), \ulambda)\in {\cal L}$,  we can use ${\cal H}_{C}(\Psi(0), \ulambda)$
in the estimate above instead of ${\cal H}_{C}(\Psi(t), \ulambda)$.
\hfill\lanbox 

\bigskip

After establishing the nonlinear stability of the stationary states of the
semi-relativistic Schr\"odinger-Poisson system, our main goal is show the
existence of such states satisfying the assumptions of the stability theorem.

\bigskip


\setcounter{equation}{0}

\section{Dual functionals}

\bigskip

For every distribution function $f\in {\cal C}$ we will derive a corresponding
stationary state as the unique maxinizer of a functional defined below. Let us
use the energy-Casimir functional from the stability result to obtain such a
dual functional. Our tool below will be the saddle point principle. Recall that, for $\Lambda>0$ fixed
$$
{\cal G}(\Psi, \ulambda,V, \sigma):=\sum_{k=1}^{\infty}[F^{*}(-\lambda_{k})+
\lambda_{k}\int_{\Omega}[|T_{m}^{\frac{1}{2}}\psi_{k}|^{2}+V|\psi_{k}|^{2}]dx]-
\frac{1}{2}\int_{\Omega}|\nabla V|^{2}dx+\sigma \Big[\sum_{k=1}^{\infty}\lambda_{k}-
\Lambda \Big].
$$
Here as before $\Psi=\{\psi_{k}\}_{k=1}^{\infty}\subset H_{0}^{\frac{1}{2}}(\Omega)
\cap H^{1}(\Omega)$ is a complete orthonormal system in $L^{2}(\Omega)$ and
$\ulambda\in l_{+}^{1}=\{(\lambda_{k})\in l^{1} \ | \ \lambda_{k}\geq 0, \ k\in
{\mathbb N} \}$. Now the function $V\in H_{0}^{1}(\Omega)$ is allowed to vary 
independently of $\Psi$ and $\ulambda$. The parameter $\sigma\in {\mathbb R}$
here plays the role of   Lagrange multipliers. The statement below demostrates
how the functional defined above is related to our energy-Casimir functional.
 
\bigskip

{\bf Lemma 9.} {\it For arbitrary $\Psi, \ulambda, \sigma$,
\begin{equation}
\label{sG}
\hbox{sup}_{V}{\cal G}(\Psi,\ulambda, V, \sigma)={\cal H}_{C}(\Psi,\ulambda)+
\sigma \Big[\sum_{k=1}^{\infty}\lambda_{k}-\Lambda \Big].
\end{equation}
The supremum is attained at $V=V_{\psi, \lambda}$.}

\bigskip

{\it Proof.} We express the functional defined above as
$$
{\cal G}(\Psi,\ulambda, V, \sigma)=\sum_{k=1}^{\infty}[F^{*}(-\lambda_{k})+
\lambda_{k}\int_{\Omega}|T_{m}^{\frac{1}{2}}\psi_{k}|^{2}dx+\frac{1}{2}\lambda_{k}
\int_{\Omega}|\psi_{k}|^{2}V_{\psi, \lambda}dx]+\sum_{k=1}^{\infty}\lambda_{k}
\int_{\Omega}V|\psi_{k}|^{2}dx-
$$
$$
-\frac{1}{2}\int_{\Omega}|\nabla V_{\psi, \lambda}|^{2}dx-
\frac{1}{2}\int_{\Omega}|\nabla V|^{2}dx+
\sigma \Big[\sum_{k=1}^{\infty}\lambda_{k}-\Lambda \Big].
$$
Using the definition of the energy-Casimir functional (\ref{HC}) we  
arrive at
$$
{\cal H}_{C}(\Psi, \ulambda)-\int_{\Omega}V\Delta V_{\psi, \lambda}dx-
\frac{1}{2}\int_{\Omega}|\nabla V_{\psi, \lambda}|^{2}dx-\frac{1}{2}\int_{\Omega}
|\nabla V|^{2}dx+\sigma \Big[\sum_{k=1}^{\infty}\lambda_{k}-\Lambda \Big].
$$
The expression above can be written as
$$
{\cal H}_{C}(\Psi, \ulambda)-\frac{1}{2}\|\nabla V_{\psi, \lambda}-\nabla V\|_
{L^{2}(\Omega)}^{2}+\sigma \Big[\sum_{k=1}^{\infty}\lambda_{k}-\Lambda \Big],
$$  
which completes the proof of the lemma.  \hfill\lanbox

\bigskip

 In the next
Section, we will show that the functional $\Phi(V,\sigma)$ defined in (\ref{fi}) has a unique 
maximizer, which is a stationary state of our Schr\"odinger-Poisson system.
Let us first prove the following auxiliary statement, which is the 
generalization of Lemma 8 above.

\bigskip

{\bf Lemma 10.} {\it Let $V\in H_{0}^{1}(\Omega)$ and $V\geq 0$. Then for 
$(\Psi,\ulambda)\in {\cal L}$ and ${\sigma\in \mathbb R}$, the lower bound}
\begin{equation}
\label{l9in}
\sum_{k=1}^{\infty}\Big[F^{*}(-\lambda_{k})+\lambda_{k}\Big(\int_{\Omega}[|T_{m}^
{\frac{1}{2}}\psi_{k}|^{2}+V|\psi_{k}|^{2}]dx+\sigma \Big)\Big]\geq -\hbox{Tr}
[F(T_{m}+V+\sigma)]
\end{equation}
{\it is valid. Equality in it is attained when 
$(\Psi,\ulambda)=(\Psi_{V},\ulambda_{V})$, where
$\psi_{V,k}\in H_{0}^{\frac{1}{2}}(\Omega)\cap H^{1}(\Omega), \ k\in 
{\mathbb N}$ is the orthonormal sequence of eigenfunctions of the operator 
$T_{m}+V$ corresponding to eigenvalues $\mu_{V,k}$. Moreover, 
$\lambda_{V,k}=f(\mu_{V,k}+\sigma), \ k\in {\mathbb N}.$}

\bigskip

{\it Proof.} Let us use inequality (\ref{Flmk}) with
$$
\mu_{k}:=\int_{\Omega}\Big(|T_{m}^{\frac{1}{2}}\psi_{k}|^{2}+V|\psi_{k}|^{2}\Big)dx+
\sigma=\langle \psi_{k}, (T_{m}+V+\sigma)\psi_{k}\rangle, \quad k\in {\mathbb N}.
$$
Therefore,
\begin{equation}
\label{F*lkF}
F^{*}(-\lambda_{k})+\lambda_{k}\Big(\int_{\Omega}\Big[|T_{m}^{\frac{1}{2}}\psi_{k}|^{2}
+V|\psi_{k}|^{2}\Big]dx+\sigma \Big)\geq -F(\langle \psi_{k}, (T_{m}+V+\sigma)
\psi_{k}\rangle), \quad k\in {\mathbb N}.
\end{equation}
Clearly,
$$
T_{m}+V+\sigma=\int_{0}^{\infty}(\lambda+\sigma)dE_{\lambda},
$$
where $E_{\lambda}$ is the spectral family associated with the Hamiltonian
$T_{m}+V$, such that 
$d\nu_{k}(\lambda):=\langle \psi_{k}, dE_{\lambda} \psi_{k}\rangle$ is a 
probability measure for $k\in {\mathbb N}$. By means of Jensen's inequality
$$
F(\langle \psi_{k}, (T_{m}+V+\sigma)\psi_{k}\rangle)=F\Big(\int_{0}^{\infty}
(\lambda+\sigma)d\nu_{k}(\lambda)\Big)\leq \int_{0}^{\infty}F(\lambda+\sigma)
d\nu_{k}(\lambda)=
$$
$$
=\langle \psi_{k}, F(T_{m}+V+\sigma)\psi_{k}\rangle.
$$
This upper bound along with (\ref{F*lkF}) and summation over 
$k\in {\mathbb N}$ give us the desired inequality (\ref{l9in}).

Then consider 
$\{\psi_{V,k}\}_{k=1}^{\infty}\subset H_{0}^{\frac{1}{2}}(\Omega)\cap H^{1}(\Omega)$
forming a complete orthonormal system in $L^{2}(\Omega)$, such that
$(T_{m}+V)\psi_{V,k}=\mu_{V,k}\psi_{V,k}$ and 
$\lambda_{V,k}=f(\mu_{V,k}+\sigma), \ k\in {\mathbb N}$. In this case the
right side of (\ref{l9in}) is equal to
$$
-\sum_{k=1}^{\infty}\langle F(T_{m}+V+\sigma)\psi_{V,k}, \psi_{V,k} \rangle=
-\sum_{k=1}^{\infty}F(\mu_{V,k}+\sigma).
$$
We have for $k\in {\mathbb N}$
$$
F^{*}(-\lambda_{V,k})=\hbox{sup}_{\lambda\in {\mathbb R}}(-\lambda \lambda_{V,k}-
F(\lambda))=-f^{-1}(\lambda_{V,k})\lambda_{V,k}-F(f^{-1}(\lambda_{V,k})),
$$
since it is attained at the maximal point $\lambda^{*}:=f^{-1}(\lambda_{V,k})$.
The equality $\lambda_{V,k}=f(\mu_{V,k}+\sigma)$ yields
$f^{-1}(\lambda_{V,k})=\mu_{V,k}+\sigma$, such that
$$
F^{*}(-\lambda_{V,k})=-(\mu_{V,k}+\sigma)\lambda_{V,k}-F(\mu_{V,k}+\sigma).
$$
A direct computation implies that the left side of (\ref{l9in}) equals to
$-\sum_{k=1}^{\infty}F(\mu_{V,k}+\sigma).$ \hfill\lanbox

\bigskip

Armed with the auxiliary lemma above we manage to derive the expression
for the dual functional for our problem.

\bigskip

{\bf Lemma 11.} {\it  The infimum in definition (\ref{fi}) is attained at
$\Psi=\{\psi_{V,k}\}_{k=1}^{\infty}$, an orthonormal sequence of eigenfunctions
of the Hamiltonian $T_{m}+V, \ V\geq 0$ corresponding to eigenvalues 
$\mu_{V,k}$ with $\lambda_{V,k}=f(\mu_{V,k}+\sigma)$ for $k\in {\mathbb N}$. 
Furthermore, the dual functional is given by}
\begin{equation}
\label{FiVs}
\Phi(V,\sigma)=-\frac{1}{2}\int_{\Omega}|\nabla V|^{2}dx-
\hbox{Tr}[F(T_{m}+V+\sigma)]-\sigma \Lambda.
\end{equation}
{\it Proof.} Let us show that the operator $F(T_{m}+V+\sigma)$ is trace class.
Clearly,
$$
\hbox{Tr}[F(T_{m}+V+\sigma)]=\sum_{k=1}^{\infty}F(\mu_{V,k}+\sigma).
$$
Since the potential function $V\geq 0$ by assumption, we use inequalities
(\ref{mkl}) and (\ref{Fu}) and arrive at the series with the general term
$(1+Ck^{\frac{1}{3}}-m+\sigma)^{-4-\varepsilon}$. This series is
convergent. We conclude the proof of the lemma by referring to the result
of Lemma 10 above. \hfill\lanbox

\bigskip


\setcounter{equation}{0}

\section{Existence of stationary states}
\label{sec:ExStatStates}

\bigskip

In this section we prove, for each distribution function $f\in {\cal C}$ and each value of 
$\Lambda>0$, the existence of a unique maximizer of the
functional $\Phi$, which will be a stationary state of the semi-relativistic
Schr\"odinger-Poisson system. 

{\it Proof of Theorem 2.} Let us first show that the bound
\begin{equation}
\label{Trb}
\hbox{Tr}[F(T_{m}+\alpha(V_{1}+\sigma_{1})+(1-\alpha)(V_{2}+\sigma_{2})]\leq
\alpha \hbox{Tr}[F(T_{m}+V_{1}+\sigma_{1})]+(1-\alpha)
\hbox{Tr}[F(T_{m}+V_{2}+\sigma_{2})]
\end{equation}
holds for any 
$\alpha\in (0,1)$ and $(V_{j},\sigma_{j})\in H_{0,+}^{1}(\Omega)\times 
{\mathbb R}, \ j=1,2$. Let $\phi\in H_{0}^{\frac{1}{2}}(\Omega)\cap H^{1}(\Omega)$
and $\|\phi\|_{L^{2}(\Omega)}=1$. We make use of the spectral decompositions
$$
T_{m}+V_{1}=\int_{0}^{\infty}\gamma dP_{\gamma}, \quad
T_{m}+V_{2}=\int_{0}^{\infty}\beta dQ_{\beta},
$$
where $P_{\gamma}$ and $Q_{\beta}$ are the spectral families associated with the
operators $T_{m}+V_{1}$ and $T_{m}+V_{2}$ respectively. This enables us to 
introduce the probability measures
\begin{equation}
\label{numu}
d\nu (\gamma):=(\phi,dP_{\gamma}\phi)_{L^{2}(\Omega)}, \quad
d\mu(\beta):=(\phi, dQ_{\beta}\phi)_{L^{2}(\Omega)}
\end{equation}
and write
$$
F((\phi, [T_{m}+\alpha(V_{1}+\sigma_{1})+(1-\alpha)(V_{2}+\sigma_{2})]\phi)
_{L^{2}(\Omega)})=
$$
$$
=F(\alpha \int_{0}^{\infty}(\gamma+\sigma_{1})d\nu(\gamma)+
(1-\alpha)\int_{0}^{\infty}(\beta+\sigma_{2})d\mu(\beta)). 
$$
Since $F$ is strictly convex on its support, we obtain the upper bound for 
the expression above using Jensen's inequality as
$$
\alpha \int_{0}^{\infty}F(\gamma+\sigma_{1})d\nu(\gamma)+(1-\alpha)\int_{0}^{\infty}
F(\beta+\sigma_{2})d\mu(\beta).
$$
By means of definition (\ref{numu}) we arrive at
$$
\alpha (\phi,F(T_{m}+V_{1}+\sigma_{1})\phi)_{L^{2}(\Omega)}+(1-\alpha)
(\phi,F(T_{m}+V_{2}+\sigma_{2})\phi)_{L^{2}(\Omega)}.
$$
Let $\{\psi_{k}\}_{k=1}^{\infty}$ be the set of eigenfunctions of the operator
$T_{m}+\alpha (V_{1}+\sigma_{1})+(1-\alpha)(V_{2}+\sigma_{2})$ forming a 
complete orthonormal system in $L^{2}(\Omega)$. Then via the argument above
we obtain
$$
\sum_{k=1}^{\infty}F((\psi_{k}, [T_{m}+\alpha(V_{1}+\sigma_{1})+(1-\alpha)
(V_{2}+\sigma_{2})]\psi_{k})_{L^{2}(\Omega)})\leq
$$
$$
\leq \alpha \sum_{k=1}^{\infty}(\psi_{k},F(T_{m}+V_{1}+\sigma_{1})\psi_{k})
_{L^{2}(\Omega)}+(1-\alpha)\sum_{k=1}^{\infty}(\psi_{k},F(T_{m}+V_{2}+\sigma_{2})
\psi_{k})_{L^{2}(\Omega)}
$$
and arrive at inequality (\ref{Trb}).
Suppose equality holds. From the fact that the function $F$ is strictly convex on its support we 
deduce that the operators $T_{m}+V_{1}+\sigma_{1}$ and $T_{m}+V_{2}+\sigma_{2}$
with potential functions $V_{1}$ and $V_{2}$ vanishing on the boundary of 
$\Omega$, have the same set of eigenvalues and the corresponding 
eigenfunctions are $\{\psi_{k}\}_{k=1}^{\infty}$. Therefore, $V_{1}(x)=V_{2}(x)$
in $\Omega$ and $\sigma_{1}=\sigma_{2},$ and  $\hbox{Tr}[F(T_{m}+V+\sigma)]$ is
strictly convex. Since $-\frac{1}{2}\int_{\Omega}|\nabla V|^{2}dx$ and 
$-\sigma \Lambda$ are concave, we obtain that our functional given by 
(\ref{FiVs}) is strictly concave.

Then we turn our attention to the proof of its boundedness from above and
coercivity. Obviously, by means of the Poincar\'e inequality
$$
\frac{1}{2}\int_{\Omega}|\nabla V|^{2}dx\geq \frac{C_{1}}{2}
\|V\|_{H_{0}^{1}(\Omega)}^{2}
$$
with a constant $C_{1}>0$. Let $\mu_{V}$ be the lowest eigenvalue of the 
Hamiltonian $T_{m}+V$. Clearly, we have the estimate with a trial function 
$\tilde{\phi}$ as
$$
\mu_{V}\leq \int_{\Omega}\{||p|^{\frac{1}{2}}\tilde{\phi}|^{2}+V|\tilde{\phi}|^{2}\}
dx, \quad \|\tilde{\phi}\|_{L^{2}(\Omega)}=1.
$$
Let us fix $\tilde{\phi}$ as the ground state of the negative Dirichlet 
Laplacian on $L^{2}(\Omega)$. Then 
$$
\int_{\Omega}||p|^{\frac{1}{2}}\tilde{\phi}|^{2}dx=\sqrt{C_{p}},
$$
where $C_{p}$ is the constant in the Poincar\'e inequality. We introduce
$$
C_{2}:=\sqrt{\int_{\Omega}|\tilde{\phi}|^{4}dx}>0,
$$
which is finite. Indeed, $\tilde{\phi}\in L^{6}(\Omega)$ via the Sobolev
inequality (\ref{sineq}). Hence via the Schwarz inequality we arrive at
$$
\int_{\Omega}V|\tilde{\phi}|^{2}dx\leq C_{2}\|V\|_{L^{2}(\Omega)}\leq C_{2}
\|V\|_{H_{0}^{1}(\Omega)},
$$
such that
$$
\mu_{V}\leq \sqrt{C_{p}}+C_{2}\|V\|_{H_{0}^{1}(\Omega)}.
$$
This yields the upper bound
\begin{equation}
\label{Fiub1}
\Phi(V,\sigma)\leq -\frac{C_{1}}{2}\|V\|_{H_{0}^{1}(\Omega)}^{2}-
F(\sqrt{C_{p}}+C_{2}\|V\|_{H_{0}^{1}(\Omega)}+\sigma)-\sigma \Lambda.
\end{equation}
Let us use the convexity property, such that
$$
F(x)\geq -\beta x+C_{3},
$$
where $\beta>\Lambda>0$ is large enough. This implies the inequality
$$
\Phi(V,\sigma)\leq -\frac{C_{1}}{2}\|V\|_{H_{0}^{1}(\Omega)}^{2}+(\beta-\Lambda)
\sigma+\beta C_{2}\|V\|_{H_{0}^{1}(\Omega)}+\beta \sqrt{C_{p}}-C_{3}.
$$
A direct computation yields the estimate
$$
\Phi(V,\sigma)\leq -\frac{C_{1}}{4}\|V\|_{H_{0}^{1}(\Omega)}^{2}+C_{5}+(\beta-\Lambda)
\sigma+\beta \sqrt{C_{p}}-C_{3}.
$$
Let us choose $\beta=2\Lambda$ and define the nonnegative constant 
$k:=\hbox{max}\{C_{5}+\beta \sqrt{C_{p}}-C_{3}, 0\}$. Hence
\begin{equation}
\label{Fiub2}
\Phi(V,\sigma)\leq -\frac{C_{1}}{4}\|V\|_{H_{0}^{1}(\Omega)}^{2}+\Lambda \sigma+k.
\end{equation}
Combining estimates (\ref{Fiub1}) and (\ref{Fiub2}), we easily arrive at
$$
\Phi(V,\sigma)\leq -\frac{C_{1}}{4}\|V\|_{H_{0}^{1}(\Omega)}^{2}-\Lambda |\sigma|+k,
$$
which shows that our functional $\Phi(V,\sigma)$ is bounded above and 
$-\Phi(V,\sigma)$ is coercive. Therefore, $\Phi(V,\sigma)$ has a unique
maximizer $(V_{0},\sigma_{0})$. Let the hamiltonian $T_{m}+V_{0}$ have the
sequence of eigenvalues $\{\mu_{0,k}\}_{k=1}^{\infty}$ and corresponding
eigenfunctions $\{\psi_{0,k}\}_{k=1}^{\infty}$, such that 
$$
(T_{m}+V_{0})\psi_{0,k}=\mu_{0,k}\psi_{0,k}, \quad k\in {\mathbb N} 
$$
and let $\lambda_{0,k}:=f(\mu_{0,k}+\sigma_{0})$. We have
$$
\Phi(V_{0},\sigma)=-\frac{1}{2}\int_{\Omega}|\nabla V_{0}|^{2}dx-\sum_{k=1}^{\infty}
\int_{\mu_{0,k}+\sigma}^{\infty}f(\xi)d\xi-\sigma \Lambda,
$$
such that $\sigma=\sigma_{0}$ is its critical point. Therefore,
$$
0=\frac{d\Phi}{d\sigma}(V_{0},\sigma)|_{\sigma=\sigma_{0}}=-\Lambda+
\sum_{k=1}^{\infty}f(\mu_{0,k}+\sigma_{0})=\sum_{k=1}^{\infty}\lambda_{0,k}-\Lambda,
$$
such that $\sum_{k=1}^{\infty}\lambda_{0,k}=\Lambda$. The first variation of 
$\Phi(V, \sigma_{0})$ at $V=V_{0}$ vanishes as well. Thus, an easy computation
gives us
$$
-\Delta V_{0}(x)=\sum_{k=1}^{\infty}\lambda_{0,k}|\psi_{0,k}(x)|^{2}.
$$
By direct substitution, the functions 
$\psi_{k}(x,t)=e^{-i\mu_{0,k}t}\psi_{0,k}(x), \ k\in {\mathbb N}$
satisfy the Schr\"odinger equation
$$
i\frac{\partial \psi_{k}}{\partial t}=[T_{m}+V_{0}]\psi_{k}, \quad x\in \Omega,
\quad t\geq 0. 
$$
The density matrix
$$ 
\rho_{0}(t,x,y)=\sum_{k=1}^{\infty}\lambda_{0,k}\psi_{k}(x,t)
\bar{\psi_{k}}(y,t)=\sum_{k=1}^{\infty}\lambda_{0,k}\psi_{0,k}(x)\psi_{0,k}(y),
$$
such that $\displaystyle{\frac{\partial \rho_{0}}{\partial t}=0}$ and the 
particle concentration $n_{0}(t,x)=\rho_{0}(t,x,x)$. 

Therefore, $(\Psi_{0}, \ulambda_{0}, \mu_{0}, V_{0})$ is a stationary state of 
our semi-relativistic Schr\"odinger-Poisson system. Finally, we are in position
to show that $(\Psi_{0}, \ulambda_{0})\in {\cal L}$, which can be done 
analogously to the proof of Lemma 6 above.  \hfill\lanbox

\bigskip

We have the following result relating the functional $\Phi$ and ${\cal H}_C.$

\bigskip

{\bf Proposition 12.} {\it Let the assumptions of Theorem 2 hold, such that
$(\Psi_{0}, \ulambda_{0}, \mu_{0}, V_{0})$ is the corresponding stationary state
of the semi-relativistic Schr\"odinger-Poisson system.  Then 
$\Phi(V_{0}, \sigma_{0})={\cal H}_{C}(\Psi_{0}, \ulambda_{0})$.}

\bigskip

{\it Proof.} Note that
$$
\Phi(V_{0}, \sigma_{0})=-\frac{1}{2}\int_{\Omega}|\nabla V_{0}|^{2}dx-
\hbox{Tr}[F(T_{m}+V_{0}+\sigma_{0})]-\sigma_{0}\Lambda
$$
and
$$
{\cal H}_{C}(\Psi_{0}, \ulambda_{0})=\sum_{k=1}^{\infty}F^{*}(-\lambda_{0,k})+
\sum_{k=1}^{\infty}\lambda_{0,k}\int_{\Omega}|T_{m}^{\frac{1}{2}}\psi_{0,k}|^{2}dx+
\frac{1}{2}\int_{\Omega}|\nabla V_{0}|^{2}dx.
$$
By means of Lemma 10
$$
\sum_{k=1}^{\infty}\Big[F^{*}(-\lambda_{0,k})+\lambda_{0,k}\Big(\int_{\Omega}[
|T_{m}^{\frac{1}{2}}\psi_{0,k}|^{2}+V_{0}|\psi_{0,k}|^{2}]dx+\sigma_{0} \Big)\Big]=
-\hbox{Tr}[F(T_{m}+V_{0}+\sigma_{0})],
$$
which implies the statement of the proposition. \hfill\lanbox

\bigskip

{\bf Acknowledgements}

\bigskip

V.V. thanks R. Jerrard and I.M. Sigal for stimulating discussions. In 
particular, the idea to pursue this problem stems from discussions and 
seminars  on related topics  initiated by I.M. Sigal that V.V. was part of, 
around 2007  at the University of Toronto. The work of T.C. was supported by 
NSF grants DMS-1009448 and DMS-1151414 (CAREER). 
W.A.S. acknowledges the support of NSERC discovery 
grant and USASK start-up fund. 

\bigskip

\section*{Appendix: Higher regularity}
In this Appendix, we extend the global well-posedness result of \cite{ACV11} to spaces with higher regularity.
For $s\in {\mathbb N},$ let
$$
{\cal L}^s:=\{ (\Psi, \ulambda) \ | \ \Psi=\{\psi_{k}\}_{k=1}^{\infty}\subset
H_{0}^{\frac{1}{2}}(\Omega)\cap H^{s}(\Omega) \; \; is \ a \ complete \ 
orthonormal \ system \ in \ L^{2}(\Omega), 
$$
$$
\ulambda=\{\lambda_{k}\}_{k=1}^{\infty}\in \ell^1, \quad \lambda_{k}\geq 0, \ k\in
{\mathbb N}, \quad \sum_{k=1}^{\infty}\lambda_{k}\int_{\Omega}|(-\Delta)^{s/2} \psi_{k}|^{2}dx<\infty \}.
$$
Note that in the state spaces defined above the Dirichlet Laplacian was
used and no additional boundary conditions were required. 
We introduce inner product $(\cdot,\cdot)_{\Zoms}$ which induces the generalized inhomogenous Sobolev norm
$$ 
\|\Phi\|_{\Zoms}:=(\sum_{k=1}^{\infty}\lambda_{k}{\|\phi_{k}\|
_{H^{s}(\Omega)}^{2}})^{\frac{1}{2}},
$$
and define the corresponding Hilbert space
$$
\Zoms:=\{\Phi=\{\phi_{k}\}_{k=1}^{\infty} \ | \ \phi_{k}\in
H_{0}^{\frac{1}{2}}(\Omega)\cap H^{s}(\Omega), \ \forall \ k\in {\mathbb N}, \ 
\|\Phi\|_{\Zoms}<\infty \}.  
$$ We also introduce the generalized homogenous Sobolev norm 
$$
\|\Phi\|_{\dZoms}:=(\sum_{k=1}^{\infty}\lambda_{k}{\|(-\Delta)^{s/2} \phi_{k}\|
_{L^{2}(\Omega)}^{2}})^{\frac{1}{2}}.
$$
Clearly, $\|\Phi\|_{\dZoms}\lesssim \|\Phi\|_{\Zoms}.$ Furthermore, it follows from the Poincar\'e inequality that $\|\Phi\|_{\Zoms}\lesssim \|\Phi\|_{\dZoms}.$ This implies the following result. 
 
\bigskip

{\bf Lemma A.1.} {\it For $\Gamma\in \Zoms$ the norms 
$\|\Gamma\|_{\Zoms}$ and $\|\Gamma\|_{\dZoms}$ are equivalent. }

\bigskip

We know from \cite{ACV11} that, for $\Psi, \; \Phi\in \Zom,$
$$\|F[\Psi]\|_{\Zom}\lesssim \|\Psi\|_{\dYom}^2 \|\Psi\|_{\Zom}$$
and that 
$$\|F[\Psi]-F[\Phi]\|_{\Zom}\lesssim (\|\Psi\|^2_{\Zom}+ \|\Phi\|_{\Zom}^2) \|\Psi-\Phi\|_{\Zom}.$$ 
We have the following inequalities in spaces of higher regularity.

\bigskip

{\bf Lemma A.2.} {\it Let $\Psi, \; \Phi\in \Zoms, \ \ s\ge 2.$ Then
$$\|F[\Psi]\|_{\Zoms}\lesssim \|\Psi\|_{{\mathcal H}_{\underline{\lambda}}^{s-1}(\Omega)}^2 \|\Psi\|_{\Zoms}$$
and that 
$$\|F[\Psi]-F[\Phi]\|_{\Zoms}\lesssim (\|\Psi\|^2_{\Zoms}+ \|\Phi\|_{\Zoms}^2) \|\Psi-\Phi\|_{\Zoms}.$$ }

\bigskip

We start by proving the first inequality. The second inequality follows using a similar analysis. 
\begin{align*}
&\|F[\Psi]\|_{\Zoms}^2\lesssim \|F[\Psi]\|_{\dZoms}^2 = \sum_{k,l\ge 0} \lambda_k\lambda_l (V[\Psi]\psi_k, (-\Delta)^s V[\Psi]\psi_l) \\
&= (-1)^s \sum_{0\le |\alpha| \le s} \sum_{k,l\ge 0} \lambda_k\lambda_l (V[\Psi]\psi_k, \partial^{2\alpha} V[\Psi] \partial^{2s-2\alpha}\psi_l) \\
&\lesssim (\sum_{0\le |\alpha| \le s} \sum_{k\ge 0} \lambda_k \|\partial^\alpha V[\Psi]\partial^{s-\alpha}\psi_k\|_{L^2(\Omega)})^2 \\
&\lesssim (\sum_{0\le |\alpha| \le s-1} \sum_{k\ge 0} \lambda_k \|\partial^\alpha V[\Psi]\|_{L^6(\Omega)}\|\partial^{s-\alpha}\psi_k\|_{L^3(\Omega)} + \sum_{k\ge 0} \lambda_k \| V[\Psi]\|_{L^\infty(\Omega)}\|\partial^{s}\psi_k\|_{L^2(\Omega)})^2\\
&\lesssim (\sum_{0\le |\alpha| \le s-1} \|\partial^\alpha V[\Psi]\|_{L^6(\Omega)}\|\Psi\|_{\dot{\mathcal H}_{\underline{\lambda}}^{s-\alpha+\frac{1}{2}}(\Omega)} +  \| V[\Psi]\|_{L^\infty(\Omega)}\|\Psi\|_{\Zoms})^2
\end{align*}
where we have used the generalized Leibnitz rule on the third line, H\"older's inequality on the fourth and fifth lines, and  the Sobolev inequality
$$\|f\|_{L^{\frac{6}{3-2p}}}\lesssim \|f\|_{H^p(\Omega)}$$ on the sixth line.
It follows from the Sobolev inequality that 
$$\|V[\Psi]\|_{L^\infty(\Omega)}^2 \lesssim \| |p|^{-1/2} n[\Psi]\|_{L^2(\Omega)}^2.$$
Furthermore
\begin{align*}
\| |p|^{-1/2} n[\Psi]\|_{L^2(\Omega)}^2 &= (n[\Psi], |p|^{-1}n[\Psi])_{L^2(\Omega)}\le \|n[\Psi]\|_{L^{3/2}(\Omega)} \||p|^{-1}n[\Psi]\|_{L^3(\Omega)} \\
&\lesssim \|\Psi\|_{\dYom}^2 \||p|^{-1/2}n[\Psi]\|_{L^2(\Omega)},
\end{align*}
where we have used H\"older's inequality in the first line, and the Sobolev inequality on the second line. 
This yields $\||p|^{-1/2}n[\Psi]\|_{L^2(\Omega)} \lesssim \|\Psi\|_{\dYom}^2,$
and hence
$$\|V[\Psi]\|_{L^\infty(\Omega)} \lesssim \|\Psi\|_{\dYom}^2.$$
We now estimate $\|\partial^\alpha V(\psi)\|_{L^6(\Omega)}, \ \ 0\le |\alpha|\le s-1.$
\begin{align*}
&\|\partial^\alpha V(\psi)\|_{L^6(\Omega)}^2 \lesssim \|n[\Psi]\|_{H^{\alpha-1}(\Omega)} \\
&\lesssim \sum_{0\le |\beta|\le |\alpha|-1} \sum_{k,l\ge 0} \lambda_k\lambda_l\|\partial^{\alpha-\beta-1}\psi_k \partial^{\beta}\psi_l\|_{L^2(\Omega)}^2\\
&\lesssim \sum_{0\le |\beta|\le |\alpha|-1} \sum_{k,l\ge 0} \lambda_k\lambda_l\|\partial^{\alpha-\beta-1}\psi_k\|_{L^6(\Omega)}^2 \|\partial^{\beta}\psi_l\|_{L^3(\Omega)}^2\\
&\le \sum_{0\le |\beta|\le |\alpha|-1} \|\Psi\|_{\dot{\mathcal H}_{\underline{\lambda}}^{\alpha-\beta}(\Omega)}^2 \|\Psi\|_{\dot{\mathcal H}_{\underline{\lambda}}^{\beta+\frac{1}{2}}(\Omega)}^2 .
\end{align*}
Combining the above inequalities yields
\begin{align*}
\|F[\Psi]\|_{\Zoms}^2&\lesssim (\sum_{0\le |\alpha|\le s-1}\; \sum_{0\le |\beta|\le |\alpha|-1} \|\Psi\|_{\dot{\mathcal H}_{\underline{\lambda}}^{\alpha-\beta}(\Omega)} \|\Psi\|_{\dot{\mathcal H}_{\underline{\lambda}}^{\beta+\frac{1}{2}}(\Omega)} \|\Psi\|_{\dot{\mathcal H}_{\underline{\lambda}}^{s-\alpha+\frac{1}{2}}(\Omega)} +  \|\Psi\|_{\dYom}^2\|\Psi\|_{\Zoms})^2\\
&\lesssim \|\Psi\|^4_{\dot{{\mathcal H}_{\underline{\lambda}}}^{s-1}(\Omega)} \|\Psi\|^2_{\dZoms},
\end{align*}
and hence
$$\|F[\Psi]\|_{\Zoms}\lesssim \|\Psi\|_{{\mathcal H}_{\underline{\lambda}}^{s-1}(\Omega)}^2 \|\Psi\|_{\Zoms}.$$ 

To prove the second inequality, note that 
\begin{align*}
\|F[\Psi]-F[\Phi]\|_{\Zoms}&=\|V[\Psi](\Psi-\Phi)+(V[\Psi]-V[\Phi])\Phi\|_
{\Zoms}\\
&\le \|V[\Psi](\Psi-\Phi)\|_{\Zoms}+\|(V[\Psi]-V[\Phi])\Phi\|_
{\Zoms}.
\end{align*}
An analysis similar to the proof of the first inequality yields the local
Lipschitz continuity and completes the proof of the lemma. \hfill\lanbox

\bigskip

Lemma A.2 together with the fact that the operator $T_{m}$ generates the group $e^{-iT_{m}t}, \ t\in {\mathbb R}$, of unitary operators implies local well-posedness in ${\mathcal L}^s.$ 
Furthermore, we know from \cite{ACV11} that, for $\Psi(x,0)\in \Zom,$ $\|\Psi(x,t)\|_{\Zom}$ is bounded for all times. 

\bigskip

{\bf Theorem A.3.}{\it For every initial state $(\Psi(x,0),\ulambda)\in {\cal L}^s, \ \ s\ge2,$
there is a unique mild solution $\Psi(x,t)$, $t\in [0,\infty)$, of 
(\ref{sch})-(\ref{n}) with $(\Psi(x,t),\ulambda)\in {\cal L}^s$, 
which is also a unique strong global solution in $\Xom$.}

The proof follows from the blow-up alternative and the first inequality in Lemma A.2. The mild solution of the Schr\"odinger-Poisson system (\ref{sch})-(\ref{n}), given by
\eqn\label{eq:mildsol-1}
	\Psi(t)=e^{-iT_m t} \Psi(0) + e^{-iT_mt} \int_0^t e^{iT_mt'}F[\Psi(t')]dt',
\eeqn
which implies
$$\|\Psi(t)\|_{\Zoms} \le \|\Psi (0)\|_{\Zoms} + \int_0^t \|F[\Psi(t')]\|_{\Zoms} dt'.$$
If $\|\Psi(t)\|_{{\mathcal H}_{\underline{\lambda}}^{s-1}(\Omega)}\lesssim 1,$ it follows from Lemma A.2 that 
$$\|\Psi(t)\|_{\Zoms} \le C_1 + C_2 \int_0^T \|\Psi(t')\|_{\Zoms} dt'.$$
By Gronwall's lemma, 
$$\|\Psi(t)\|_{\Zoms} \le C_1e^{C_2 t}, \ \ t\in [0,T']\subset [0,T).$$
Since $\|\Psi(t)\|_{\Zom}$ is bounded for all times, it follows by induction on $s$ that $\|\Psi(t)\|_{\Zoms}$ is bounded for all times. \hfill\lanbox

\end{document}